\documentclass[10pt,twocolumn,letterpaper]{article}

\usepackage{iccv}
\usepackage{times}
\usepackage{epsfig}
\usepackage{graphicx}
\usepackage{amsmath}
\usepackage{amssymb}
\usepackage{multirow}
\usepackage{stfloats}

\usepackage{color, colortbl}
\usepackage{makecell}
\definecolor{LightCyan}{rgb}{0.88,1,1}

\usepackage[pagebackref=true,breaklinks=true,letterpaper=true,colorlinks,bookmarks=false]{hyperref}

\iccvfinalcopy 

\def\iccvPaperID{8977} 

\ificcvfinal\fi

\begin{document}

\title{HCiM: ADC-Less \underline{H}ybrid Analog-Digital \underline{C}ompute \underline{i}n \underline{M}emory Accelerator for Deep Learning Workloads}

\author{Shubham Negi, Utkarsh Saxena, Deepika Sharma, and Kaushik Roy\\
Elmore Family School of Electrical and Computer Engineering, Purdue University\\
West Lafayette, IN 47907, USA\\
{\tt\small snegi@purdue.edu}
}

\maketitle
\ificcvfinal\thispagestyle{empty}\fi

\newcommand{\papername}{HCiM }
\newcommand{\papernamewospace}{HCiM}

\begin{abstract}
Analog Compute-in-Memory (CiM) accelerators are increasingly recognized for their efficiency in accelerating Deep Neural Networks (DNN). However, their dependence on Analog-to-Digital Converters (ADCs) for accumulating partial sums from crossbars leads to substantial power and area overhead. Moreover, the high area overhead of ADCs constrains the throughput due to the limited number of ADCs that can be integrated per crossbar. An approach to mitigate this issue involves the adoption of extreme low-precision quantization (binary or ternary) for partial sums. Training based on such an approach eliminates the need for ADCs. While this strategy effectively reduces ADC costs, it introduces the challenge of managing numerous floating-point scale factors, which are trainable parameters like DNN weights. These scale factors must be multiplied with the binary or ternary outputs at the columns of the crossbar to ensure system accuracy. To that effect, we propose an algorithm-hardware co-design approach, where DNNs are first trained with quantization-aware training. Subsequently, we introduce \papername, an ADC-Less Hybrid Analog-Digital CiM accelerator. \papername uses analog CiM crossbars for performing Matrix-Vector Multiplication operations coupled with a digital CiM array dedicated to processing scale factors. This digital CiM array can execute both addition and subtraction operations within the memory array, thus enhancing processing speed. Additionally, it exploits the inherent sparsity in ternary quantization to achieve further energy savings. Compared to an analog CiM baseline architecture using 7 and 4-bit ADC, \papername achieves energy reductions up to 28$\times$ and 12$\times$, respectively.
\end{abstract}

\section{Introduction}

Machine Learning (ML) models are ubiquitous, being used everywhere from smartphones to data centers \cite{dean2022golden}. This advancement has been driven by traditional hardware such as CPUs, GPUs, and specialized TPUs, which provide the computational power needed for increasingly complex tasks \cite{hu2022survey}. However, the traditional hardware suffers from an efficiency barrier -- the "memory wall" -- arising from constant data movement between memory and processing units. This bottleneck not only slows down the computation but also increases power consumption \cite{gholami2021ai}. In response to this challenge, Compute-in-Memory (CiM) accelerators have emerged as an innovative solution, by performing computation within the memory array itself, thereby reducing data movement \cite{ankit2020circuits}.

\begin{figure}[t]
\centering
\includegraphics[width=0.4\textwidth]{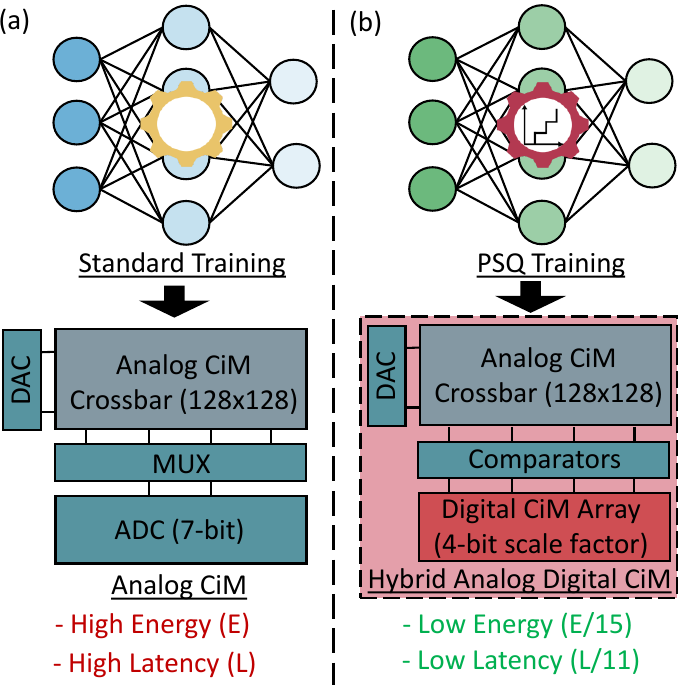}
    \caption{(a) ResNet-20 trained with standard training mapped to CiM hardware. (b) ResNet-20 trained with Partial-Sum Quantization (PSQ) training mapped to \papername has $15\times$ and $11\times$ lower energy and latency. Area-normalized latency is reported, reflecting differences in baseline areas.}
\centering
\label{summary}
\vskip -0.2 in
\end{figure}

While CiM accelerators are a promising candidate for overcoming traditional hardware limitations, they introduce a new challenge associated with Analog-to-Digital Converters (ADCs). ADCs are essential for accumulating partial sum outputs from multiple crossbar arrays \cite{puma}, and are a primary source of power and area inefficiency within the CiM architecture. Note, ADCs alone consume as much as 60\% energy and occupy nearly 80\% area in CiM accelerators \cite{nax}. Furthermore, due to the high area requirement of the ADCs, they are shared across multiple columns in the crossbar, thereby limiting the throughput of the crossbar arrays.

Researchers have explored several strategies to mitigate the limitations imposed by ADCs in CiM accelerators \cite{kim2022extreme, bitsplitnet, quarry, islped}. Many of these works have focused on scaling the precision of ADCs to lower bit levels by Partial Sum Quantization (PSQ), to save on power and area. Yet, even with these lower precision ADCs, the challenge of inefficiency remains. Advancing on the PSQ technique, recent work has shown that the partial sums can be quantized to just 1 (binary) or 1.5-bits (ternary) \cite{islped}. This method eliminates the need for ADCs by adopting learned step quantization \cite{lsq}. At the core of this method are scale factors, trainable parameters similar to neural network weights, which are multiplied by these binary or ternary values to align the dynamic range of quantized data with that of actual floating point data. However, processing these scale factors in hardware introduces several challenges: i) Managing a substantial number of scale factors due to the fine granularity of PSQ introduces energy and area overhead, ii) Further, these scale factors have floating-point values that require complex hardware, iii) Processing scale factors requires energy-intensive operations like multiplication or addition/subtraction, iv) Reducing the number of scale factors results in a significant drop in accuracy. Consequently, efficient processing of these scale factors is vital for the system's performance and accuracy.

To address these challenges, we present \papername, an ADC-Less Hybrid Analog-Digital CiM accelerator, to perform Matrix-Vector Multiplication (MVM) operation efficiently. It consists of an analog CiM crossbar to perform the MVM operation between inputs and weights: the partial sum outputs from the crossbar array are first digitized using a comparator and then multiplied with a scale factor. A quantization-aware training methodology quantizes the scale factors to a few bits. Finally, a digital CiM (DCiM) array is proposed to process these scale factors. Fig.~\ref{summary} compares the performance of a Deep Neural Network (DNN) deployed on conventional analog CiM hardware and a PSQ-trained neural network deployed on \papername. Our main contributions are as follows:

\begin{itemize}

    \item We introduce an algorithm-hardware co-design approach to design an ADC-Less CiM accelerator (\papername). This includes scale factor quantization-aware training and developing a Digital CiM array (to process scale factors) with a novel in-memory subtraction technique (section~\ref{codesign}).

    \item We utilize the inherent sparsity in ternary quantized partial sums, achieving a 25\% reduction in energy through clock gating in the proposed DCiM peripherals (section~\ref{sec42},~\ref{system}).

    \item We evaluate \papername using a cycle-accurate simulator across multiple workloads, demonstrating its efficacy against baselines (section~\ref{system}).
\end{itemize}

\begin{figure}[!tb]
\centering
\includegraphics[width=0.45\textwidth]{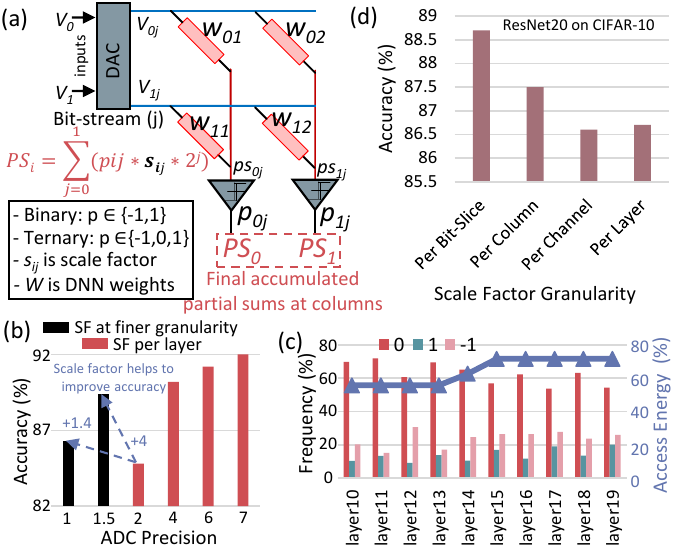}
\caption{(a) Overview of PSQ training algorithm. (b) ResNet20 accuracy with ADC precision. (c) Distribution of \textit{p} at the columns of the crossbar. Scale factor (SF) access energy compared to total off-chip data access energy. (d) Impact of reducing the number of scale factors on application accuracy.}
\centering
\label{psq}
\end{figure}

\section{Background and Challenges}\label{background}

\textbf{Analog CiM Accelerator:} Analog CiM accelerators perform MVM operations following a weight-stationary dataflow \cite{puma}. In this approach, the weights of a DNN are stored in the memory array in a bit-sliced fashion. The inputs are then bit-streamed over multiple cycles after a digital-to-analog conversion (DAC). The output at the columns of the crossbar, for each input bit stream, is converted to digital signals using an ADC. Finally, the outputs from various crossbars are accumulated after a shift and add operation. Note that bit-slice is the number of weight bits a memory cell can store and bit-stream is the DAC precision. 

\noindent\textbf{PSQ Training:} 
Partial-sum quantization (PSQ) employs a quantization-aware training approach to reduce the precision of partial-sums and hence ADC precision \cite{kim2022extreme, bitsplitnet, quarry, islped}. Notably, \cite{islped} demonstrated that partial-sum precision can be reduced to just binary or ternary values as illustrated in Fig.~\ref{psq}(a). This approach employs trainable floating-point scale factors and eliminates the need for ADCs. The binary/ternary value (\textit{p}) at the columns of crossbars are multiplied by a scale factor (\textit{s}) to bring the quantized values to a similar range as the actual floating-point data. The value of $p$ for binary/ternary quantization is shown in Eq.~\ref{eq}, where $\alpha$ is a trainable threshold.

\begin{equation}\label{eq}
\begin{split}
      p_{b} =  
    \begin{cases}
      1 & \text{if $ps \geq 0$}\\
      -1 & \text{if $ps < 0$}\\
    \end{cases}     
\end{split}
\hspace{0.4cm} \vline \hspace{0.4cm}
\begin{split}
   p_{t} = 
    \begin{cases}
      1 & \text{if $ps \geq \alpha $}\\
      0 & \text{$-\alpha < ps < \alpha $}\\
      -1 & \text{if $ps \leq -\alpha$}\\
    \end{cases}     
\end{split}
\end{equation}

\noindent To get the final partial sum value ($PS$) at the column of the crossbar, the $p*s*2^j$ ($2^j$ is for shift operation) is accumulated over all the input bit streams as shown in Fig.~\ref{background}(a). Our experiments demonstrate the necessity of using scaling factors in PSQ. Fig. \ref{psq}(b) shows that binary and ternary PSQ with scaling factors achieves higher accuracy than 2-bit quantized partial-sums.

\noindent\textbf{Challenges:} \textcircled{\textbf{1}} Number of scale factors (bit-slice granularity) per analog crossbar is shown in Eq~\ref{eq1}. Consequently, a high number of scale factors per layer can lead to significant data movement energy as depicted in Fig.~\ref{psq}(c), potentially diminishing the benefits of CiM accelerators in reducing data movement energy. One potential solution is to reduce the number of scale factors. However, as shown in Fig.~\ref{psq}(d), decreasing the number of scale factors results in reduced accuracy \cite{islped}. \textcircled{\textbf{2}} Moreover, authors in \cite{islped} use floating-point scale factors and implementing these in hardware can be energy intensive. \textcircled{\textbf{3}} Depending on the binary/ternary values of $p$, the scale factor is either subtracted from or added to the partial sum $PS$, as illustrated in Fig.~\ref{psq}(a). This requires the hardware to either support multiplication or both addition/subtraction operations. \textcircled{\textbf{4}} Additionally, in the case of ternary PSQ, many $p$ values can be zero, as seen in Fig.~\ref{psq}(c). This allows for these particular scale factor computations to be skipped. In response to these challenges, we first use quantization-aware training to quantize these scale factors, next we propose a hybrid analog-digital CiM macro that performs MVM operation in the analog CiM crossbar and the scale factor computations in digital CiM array, reducing the scale factor data movement overhead.

\begin{equation}\label{eq1}
    \text{\# Scale Factors per crossbar} = \frac{\text{input\_precision}}{\text{bit-stream}} * \text{\#columns}
\end{equation}

\section{Related Works}

Past research in the field of Analog CiM accelerators has focused on reducing ADC accesses \cite{rimac} or completely eliminating them \cite{neurocim, reram_adcless, bitsplitnet}. For instance, authors in \cite{rimac} introduced analog caches combined with analog accumulators to store the partial sums from multiple crossbars in the analog domain. However, this approach exacerbates errors due to the inherent noise in analog MVM computations. \cite{neurocim} also uses analog accumulation to remove the ADC from the CiM accelerator, however, the implementation is specifically for Spiking Neural Networks. Researchers in \cite{reram_adcless} propose a 3T2R RRAM macro, utilizing voltage division quantization for Binary DNNs. Authors in \cite{bitsplitnet} use a 1-bit sense amplifier (SA) to replace ADCs. However, this work does not consider a trainable scale factor at fine granularity. Therefore, it leads to a drop in accuracy and DNNs quantized with fine granularity of scale factors cannot be mapped to this hardware. In contrast to these works, our approach is more generic, can support any input and weight precision, and does not depend on specific memory cells in analog CiM crossbars. Moreover, since scale factors are processed in digital CiM, they do not incur any computation error.

\section{Algorithm Hardware Co-design}\label{codesign}
\subsection{Quantization aware Training}\label{training}
We adopted the PSQ training technique as described in \cite{islped}, incorporating some modifications. The original technique in \cite{islped} uses floating-point scale factors, but we recognize that managing a large quantity of these scale factors can result in substantial energy consumption and require complex floating-point hardware. To address this, along with weights, activations, and partial sums, we also quantize the scale factors into fixed-point representations at the layer level during the training phase. Quantization of scale factors introduces a single additional scale factor per layer similar to weight quantization which can be merged into the batch normalization layer. Moreover, authors in \cite{islped} use threshold ($\alpha$) from Eq.~\ref{eq} at the bit-slice level. However, this is not practically feasible therefore we use a per-layer level $\alpha$.

\begin{figure*}[t]
\centering
\includegraphics[width=0.95\textwidth]{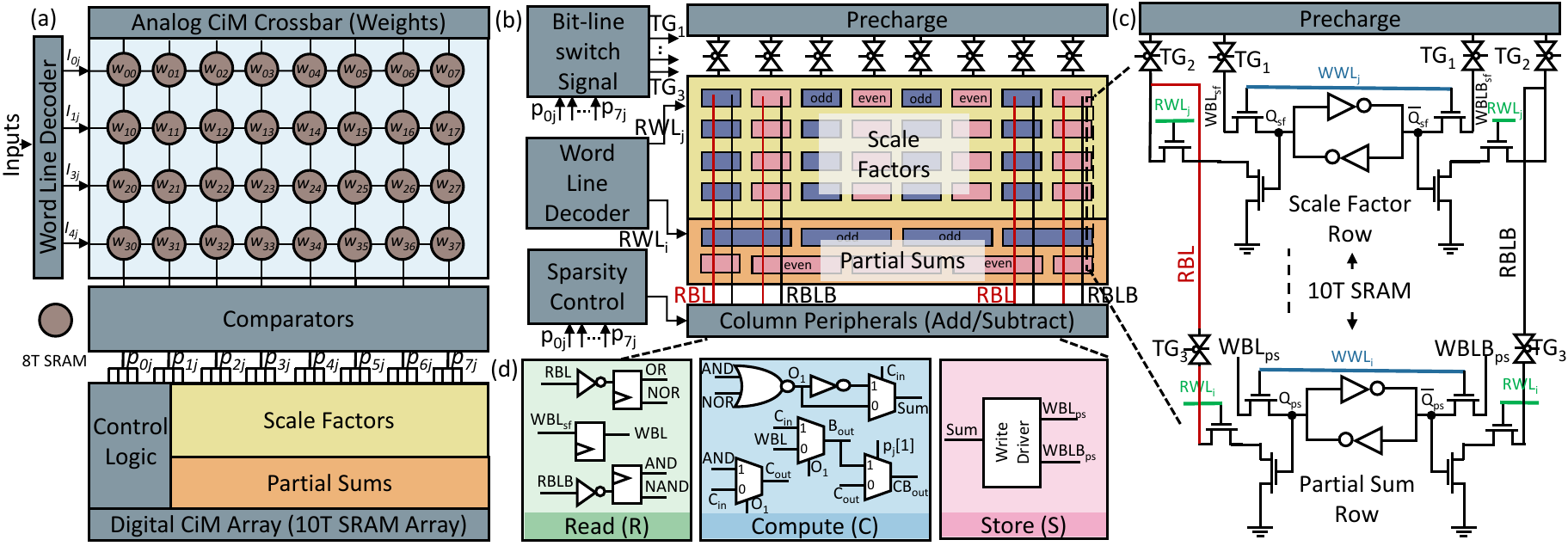}
\caption{(a) Hybrid Analog-Digital CiM macro. (b) Architecture of digital CiM array, incorporating column peripherals with a chain of 1-bit adder/subtractor at each column to enable full-adder/subtractor functionality. (c) Read Bit lines (RBL, RBLB) of scale factor and partial sum memories are connected to realize CiM operation. Write Bit line ($WBL_{sf}$) of scale factor memory helps to perform both read and write operations. (d) Detailed view of the column peripheral, illustrating the implementation of in-memory subtraction and addition operations. $CB_{out}$ represents the final carry/borrow output from the column peripheral.}
\centering
\label{architecture}
\end{figure*}

\subsection{HCiM Architecture}\label{main_arch}


The overall architecture of \papername macro is depicted in Fig.~\ref{architecture} (a). It includes an analog CiM crossbar for executing MVM operations between inputs and weights. The crossbar utilizes a charge-based 8T-SRAM \cite{cicc} to store weight data, while input activations are fed through the word line decoder. These input bits are streamed over multiple cycles. For each input bit, the column output is processed through a comparator. We require one comparator in the case of binary PSQ and two comparators in the case of ternary PSQ \cite{islped}. The output from the comparator is a set of binary/ternary values $p$(-1,0,1). Given that $p$ can take a negative value, we represent it using 2-bit numbers: 00 for 0, 01 for 1, and 11 for -1.

The quantized scale factors corresponding to a PSQ-trained network are stored in a Digital CiM (DCiM) array. The outputs from the comparator are fed into this DCiM array, which performs the scale factor computation from Fig.~\ref{psq} (a). The shift operation ($2^j$) is merged with the scale factor values during training. While the DCiM array draws inspiration from \cite{impulse}, this prior work was limited to vector-vector addition operations. However, for effective scale factor computation ($s*p$), our system requires the capability to perform addition, subtraction, or no operation, depending on whether $p$ is 1, -1, or 0. Thus, a significant contribution of our work is the adaptation of the DCiM array to efficiently execute these specific operations. The details of the DCiM array are discussed below.

\subsubsection{CiM Full Subtractor} \label{sec41}
\begin{figure}[htb]
\centering
\includegraphics[width=0.4\textwidth]{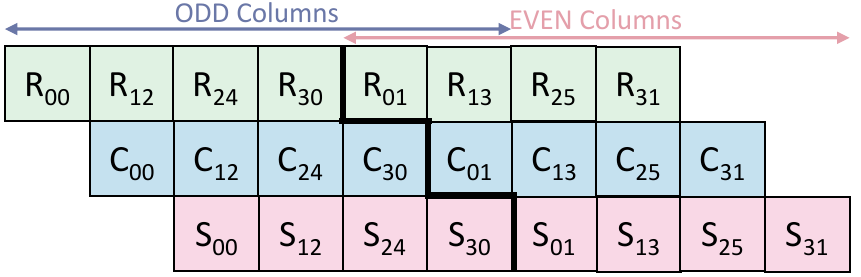}
\caption{Read Compute Store pipeline of DCiM array. $R_{ji}$ represents the addition or subtraction operation between row j and i of scale factor and partial sum memory.}
\centering
\label{pipeline}
\end{figure}

To understand the implementation of the subtraction operation, we first illustrate how an addition operation can be performed in the DCiM array. Given that the precision of partial sums is higher than that of scale factors, it requires 2 cycles to add a scale factor row to a partial sum row \cite{impulse}. This process involves handling odd columns in one cycle and even columns in the subsequent cycle. To enhance throughput, the addition operation is pipelined into 3 cycles as depicted in Fig.~\ref{pipeline}. Firstly, to add row j of the scale factor memory to row i of the partial sum memory, $RWL_{j,i}$ are activated. Next, binary/ternary values $p$ are used in the bit-line switch signal block (Fig.~\ref{architecture} (b), (c)) to generate transmission gate ($TG_{2,3}$) signals. For instance, $p$=00 turns $TG_{1,2,3}$ OFF, whereas $p$=01 turns $TG_{2,3}$ ON. As a result, on the bit lines, we get the OR and NAND of the enabled RWL, which are latched during the read cycle. This helps to process multiple columns of analog CiM crossbar in parallel. Secondly, these latched values are utilized by the column peripherals to calculate Sum and Carry ($C_{out}$) bits in the Compute cycle (Fig.~\ref{architecture} (d)). Finally, the Sum output from the compute block is stored in the partial sum memory in the Store cycle. 


A straightforward way to implement a subtraction operation with the above addition operation would be to store 2's complement of scale factors and retrieve these values whenever subtraction needs to be performed. However, this would require 2$\times$ memory since the same scale factor value can be added to or subtracted from the partial sum value depending on $p$. The logical expression of Difference (D) and Borrow ($B_{out}$) bit of a full subtractor is shown in Eq.~\ref{eq2},~\ref{eq3}. While the Difference bit is same as the Sum bit of a full adder, realizing the Borrow output using OR/NOR and NAND/AND operations is not possible. Previous works, such as in \cite{subtractor}, employed a two-cycle process for in-memory subtraction, reading one input in the first cycle and performing a bitwise operation between two inputs in the next. This approach, however, increases the latency of subtraction (4 cycles). To overcome this, our proposed architecture achieves subtraction in 3 cycles similar to an addition operation. During the read cycle, when the write bit lines and word lines of the scale factor memory are idle (as seen in Figure~\ref{architecture}(c)), we read the scale factor value in parallel for columns requiring subtraction ($p$=11) by activating $TG_{1}$. This value is then used in the Compute cycle to realize the $B_{out}$ bit, as shown in Fig.~\ref{architecture} (d). Finally, the binary/ternary value $p$ acts as a select bit to MUX in Fig.~\ref{architecture} (d) to decide if the output is a carry or borrow bit.

\begin{equation}\label{eq2}
     D = A \oplus B \oplus B_{in},\quad\text{A,B, $B_{in}$ are two inputs and input borrow}
\end{equation}
\begin{equation}\label{eq3}
     B_{out} = \overline{A}B + BB_{in} + B_{in}\overline{A}
\end{equation}

\subsubsection{Sparsity Control} \label{sec42}

As illustrated in Fig.~\ref{psq} (c) at least 50\% of ternary values $p$ are zero. Therefore, a significant portion of scale factor computation ($p*s$) can be skipped. During the Read cycle, if a particular $p$ value is zero, the bit-line switch signal deactivates that column by keeping $TG_{1,2,3}$ OFF. This prevents bit-lines from discharging, thereby saving dynamic energy. Additionally, we implement clock gating in the column peripherals, allowing us to bypass computations for columns with $p$=0 in the Compute cycle. The sparsity control block is responsible for generating the enable signal for this clock gating logic. Also, in the Store cycle, no value is written to the columns where $p$ = 0, further saving on write energy.

\begin{table*}[!htb]
    \centering    
    \caption{\papername configurations for 4-bit weight and activations. \#Scale factor and \#Partial sum per crossbar (Eq.~\ref{eq1}).}
    \label{config}
    \begin{tabular}{|c|c|c|c|c|}
        \hline
        \textbf{Configuration} & \textbf{Analog CiM Crossbar Size} & \textbf{\#Scale Factors} & \textbf{\#PartialSums} & \textbf{DCiM Array Size}\\
        
        \hline\hline    
         A & 128x128 & 4*128 & 1*128 & 24x128\\
         \hline
         B & 64x64& 4*64 & 1*64 & 24x64\\
         \hline
    \end{tabular}
\end{table*}

\section{Evaluation}\label{results}
\subsection{Experimental Setup}\label{setup}

\textbf{Training:} To demonstrate the effectiveness of our approach with quantized scale factors, we trained ResNet-20 and Wide-ResNet-20 models on the CIFAR-10 dataset and ResNet-18 model on the ImageNet dataset. The training was conducted in PyTorch using the training methodology from section~\ref{training}. For the CIFAR-10 dataset, we utilized inputs, weights and scale factors at 4-bits each, with partial sums at 8-bits. On the other hand, for the ImageNet dataset, we employed inputs and weights at 3-bits, scale factors at 8-bits, and partial sums at 16-bits.

\begin{table*}[htb]
    \centering    
    \caption{Accuracy with varying ADC precision and crossbar sizes. ADC precision 1 and 1.5 represent models trained with binary and ternary PSQ.}    
    \label{table2}
    \begin{tabular}{|c|c|c|c|c|c|}
        \hline
        \textbf{Model} & \multicolumn{5}{c|}{\textbf{ADC Precision (bits)}}\\
        \cline{2-6}
         \textbf{(Crossbar Size)} & \textbf{7} & \textbf{6} & \textbf{4} & \textbf{1.5} & \textbf{1}\\
         \hline\hline
         ResNet-20 (128) & 92.26 & 91.27 & 90.20 & 88.80& 86.30\\
         \hline
         ResNet-20 (64) & - & 91.93 & 91.00 & 89.80 & 88.20\\
         \hline
         Wide ResNet-20 (128) & 93.80  & 93.70 & 92.90 & 92.03 & 91.90\\
         \hline
         Wide ResNet-20 (64) & - & 93.91 & 93.10 & 92.24 & 91.89\\   
         \hline
   \end{tabular}
\end{table*}

\noindent\textbf{\papername:} The DCiM array, illustrated in Fig.~\ref{architecture} (b) is designed in 65nm technology. The energy, latency, and area results for the DCiM array are based on schematic-level simulation. The 10T SRAM arrays were designed with Cadence Virtuoso and the control logic with Synopsis Design Compiler. The supply voltage is set to 1V, and the operating frequency is 500 MHz. We compare the macro-level results of our DCiM array with various types of ADCs \cite{adc1, adc2, adc3} for processing all the columns of the analog CiM crossbar. To ensure a fair comparison, we selected ADCs designed in 65nm as per the ADC survey \cite{adc_survey}. For system-level comparisons, we use the cycle-accurate simulator from PUMA \cite{puma} where we replace the ADCs with our DCiM array. Considering the 32nm technology of the other components in \cite{puma}, we scaled the metrics of ADCs and our DCiM array to 32nm using predictive technology models \cite{scaling}. The energy, latency and area for the analog CiM crossbar, based on 8T SRAM cells, are derived from \cite{cicc}, with both bit-slice and bit-stream set to 1. The area for the comparators in ternary PSQ implementation is adopted from \cite{bindra20181}. Consistent with prior CiM works \cite{puma, peng2019dnn} we assume weights, once trained, are pre-loaded to the memory arrays and can be reused by different inputs. Similarly, scale factors are also pre-loaded into the memory array in line with the CiM approach, enabling their reuse across various inputs. We present accuracy and performance results for two configurations of \papername shown in Table~\ref{config}. The memory sizes for scale factor and partial sum in Configuration A are 4*128*4 and 1*128*8 bits (Table~\ref{config}). Therefore, we need a 24x128 DCiM array per analog crossbar. For benchmarking \papername we show results on CIFAR-10 dataset using a variety of models like ResNet-20,32,44 \cite{he2016deep}, Wide ResNet-20 \cite{islped}, and VGG-9,11 \cite{vgg}. Additionally, we compare \papername with previous hardware-algorithm co-design approaches \cite{bitsplitnet, quarry} aimed at reducing ADC precision on the ImageNet dataset.

\subsection{Accuracy Results}\label{accuracy}
The accuracy of neural networks trained with PSQ training is presented in Table~\ref{table2}. We observe a minimal drop in accuracy with the reduction of ADC precision, attributable to quantization-aware training. Interestingly, when partial sums are quantized to ternary values, the accuracy is comparable (within 1.5\%) to that achieved with 4-bit ADC precision. This can be credited to the fine granularity of scale factors, which effectively enhances the representation ability of the quantized partial sums. The accuracy drops $\sim$2\% as we go from 1.5 to 1-bit ADC in the ResNet-20 model which is due to the extreme quantization of partial sums. However, the drop in accuracy when moving from 1.5-bit to 1-bit ADC is less than 0.5\% for Wide Resnet-20 model. Additionally, we present results using a 64x64 analog CiM crossbar, which ideally requires only 6-bit ADCs. It is observed that the accuracy drop with 1,1.5-bit ADC is less pronounced compared to a 128x128 crossbar. This is because, in a 64x64 crossbar, the partial sums undergo less severe quantization from only 6 to 1.5-bits.

\begin{figure}[htb]
\centering
\includegraphics[width=0.45\textwidth]{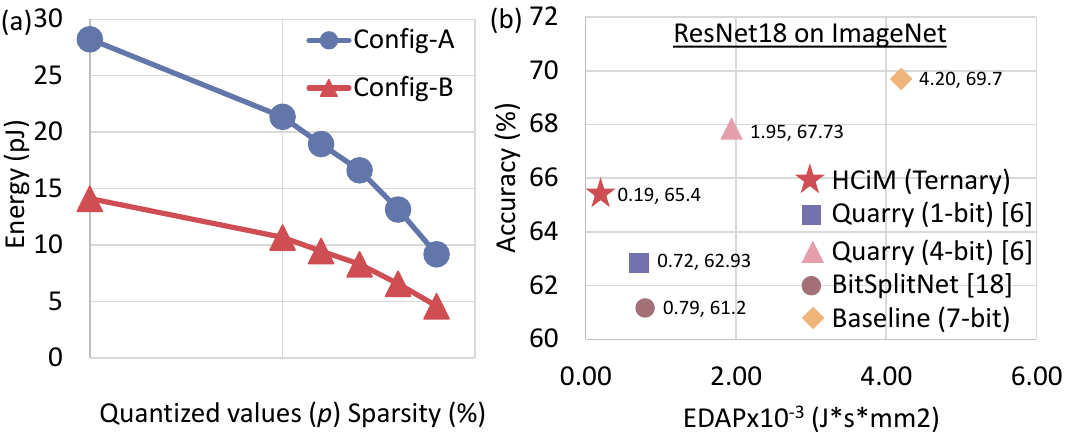}
\caption{ (a) Energy to process all the columns of analog CiM crossbar with ternary quantization. (b) Accuracy vs EDAP comparison of HCiM with baselines on ImageNet dataset.}
\centering
\label{sparsity}
\end{figure}

\begin{figure*}[htb]
\centering
\includegraphics[width=0.95\textwidth]{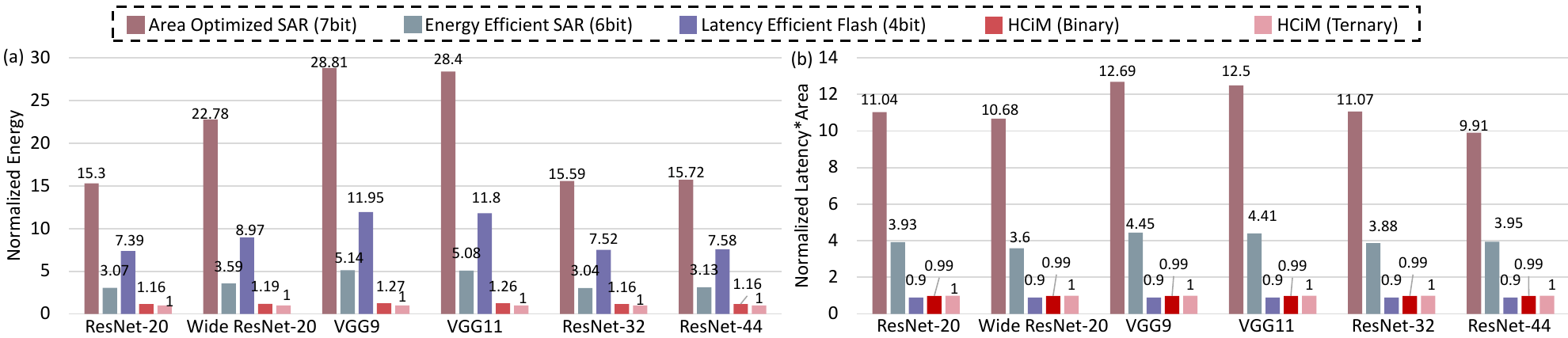}
\caption{(a) Energy and (b) Latency*Area for different workloads using \papername Configuration A from Table~\ref{config} compared to baseline using low precision ADCs. Both Energy and Latency*Area are normalized with respect to \papername (Ternary).}
\centering
\label{res1}
\vskip -0.1 in
\end{figure*}

\subsection{Performance Results}\label{system}

\noindent\textbf{Energy vs Sparsity: } The scale factor computation corresponding to the columns of analog CiM crossbar where $p$=0 can be skipped, leading to a significant reduction in energy consumption as illustrated in Fig.~\ref{sparsity} (a). Specifically going from 0\% sparsity (binary quantization) to 50\% sparsity results in 24\% energy reduction. This energy reduction is due to several factors: there is no precharge of bit-lines for these columns, adder/subtractor circuits are clock-gated, and no store operation is executed in the store cycle. However, it is crucial to note that while sparsity contributes to energy savings, it does not impact latency. This is because, in each cycle, multiple columns are processed in parallel; therefore, even with a 50\% sparsity, there may still be columns with non-zero $p$ values.

\begin{table*}[htb]
    \centering    
    \caption{DCiM array comparison with ADC to process one column of analog CiM crossbar. A \& B are configs from Table~\ref{config}}
    \label{macro_comparison}
    \begin{tabular}{|c|c|c|c|c|}
        \hline
        \textbf{\thead{Analog CiM\\ Column Peripheral}} & \textbf{\thead{ADC\\ Precision}} & \textbf{\thead{Latency\\ (ns)}} & \textbf{\thead{Energy\\ (pJ)}} & \textbf{\thead{Area\\ ($mm^2$)}}\\
        \hline\hline    
         Area Optimized SAR \cite{adc1} & 7 & 1.52 & 4.1 & 0.004\\
         \hline
         Energy Efficient SAR \cite{adc2} & 6& 0.15 & 0.59 & 0.027\\
         \hline
         Latency Efficient Flash \cite{adc3} &  4 & 0.05 & 1.86 & 0.003\\
         \hline
         DCiM Array (A) &  - & 0.06 & 0.22 & 0.009\\
         \hline
         DCiM Array (B) &  - & 0.1 & 0.22 & 0.005\\
         \hline
    \end{tabular}
\end{table*}

\noindent\textbf{DCiM array vs ADCs: } Table~\ref{macro_comparison} shows the comparison of DCiM array in \papername to different types of ADCs. Since the DCiM array can process multiple columns in parallel the latency is averaged over columns. Compared to 6,7-bit ADCs DCiM(A,B) has very low latency this is because it can process multiple columns of analog CiM crossbar in parallel. In terms of energy, DCiM(A,B) has 12$\times$ lower energy than the 4-bit ADC. DCiM (A) has 3.6$\times$ higher area normalized latency\footnote{area normalized latency (latency*area) is compared to reflect differences in area} compared to 4-bit ADC. Moreover, comparing the latency of DCiM (A) with DCiM (B) we can see that configuration A has ~2$\times$ lower latency since it can process 2$\times$ columns in parallel. For all the system level simulations in next section, we consider only 1 ADC and DCiM array per analog CiM crossbar.

\begin{figure*}[t]
\centering
\includegraphics[width=0.95\textwidth]{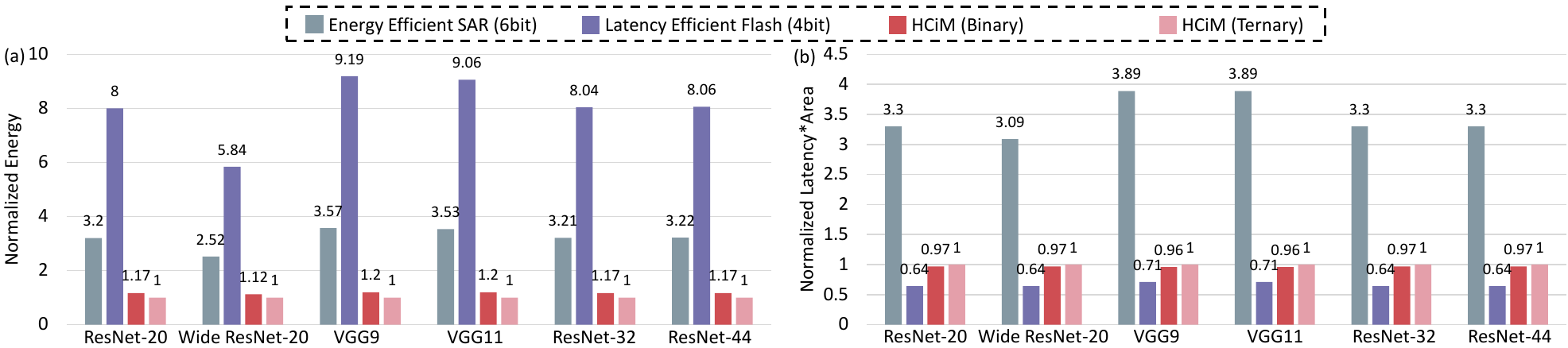}
\caption{(a) Energy and (b) Latency*Area for different workloads using \papername Configuration B from Table~\ref{config} compared to baseline using low precision ADCs. Both Energy and Latency*Area are normalized with respect to \papername (Ternary).}

\centering
\label{res2}
\vskip -0.1 in
\end{figure*}

\noindent\textbf{System Level Comparison: } \papername is compared to Analog CiM accelerators that use low precision ADCs. From Fig.~\ref{res1} (a) we can see that compared to baseline analog CiM accelerator using low precision ADCs both \papername (Binary, Ternary) have lower energy. On average across all the models \papername has at least 3$\times$ lower energy compared to all the baselines. Moreover, compared to \papername (Binary),  \papername (Ternary) has at least 15\% lower energy which is due to the sparsity support in \papername. On the other hand, \papername has at least $3-12\times$ low latency compared to baselines using SAR ADCs. This can be attributed to the low energy and latency of DCiM compared to SAR ADCs in Table~\ref{macro_comparison}. However, compared to the analog CiM baseline using 4-bit Flash ADC, \papername has 11\% higher latency. This is due to the very low latency and area of the Flash ADC \cite{adc3} as can be seen in Table~\ref{macro_comparison}.

Next, we show system-level results considering \papername configuration B where we consider 64x64 crossbar size (Fig.~\ref{res2}). Note, as we reduce the crossbar size in PUMA \cite{puma} we increase the number of crossbars so that it has the same number of multiplication units as the 128x128 crossbar size. However, increasing the number of analog crossbars would increase the data movement of partial sums across the crossbars compared to configuration A. Therefore, the advantage of reducing the ADC overhead might decrease. However, even in this case, we observe that \papername has atleast 2.5$\times$ lower energy compared to baselines that use 6-bit and 4-bit ADCs. On the other hand \papername has 1.4$\times$ higher latency compared to 4-bit ADC which is due to the low area and latency overhead of Flash ADC.

\noindent\textbf{\papername vs Related works:} The accuracy vs energy-delay-area-product (EDAP) for \papername compared to related works is shown in Fig.~\ref{sparsity}(b). Quarry \cite{quarry} and BitSplitNet \cite{bitsplitnet} did not report the performance results, we get their performance results using cycle-accurate simulator in \cite{puma}. Quarry uses analog and digital multipliers to implement scale factors. The energy and area for 1-bit ADC in \cite{quarry} is estimated as 1/16 of 4-bit flash ADC \cite{adc3}. The energy for the digital multiplier is obtained from \cite{puma}. Therefore, compared to Quarry with 1-bit ADC, \papername achieves 2.5\% higher accuracy with 3.8$\times$ lower EDAP. Moreover, compared to Quarry (4-bit), \papername has 2.3\% lower accuracy with 10.4$\times$ lower EDAP. Compared to BitSplitNet, \papername has 4.2\% higher accuracy with 4.2$\times$ lower EDAP. BitSplitNet uses independent paths to process each input and weight bit. Hence, energy and area for ResNet-18 with 4-bit inputs and weights are obtained by scaling 1-bit energy and area by 4.

\section{Conclusion}

In this work, we show how the cost of ADCs in analog CiM accelerators can be significantly reduced through an innovative algorithm-hardware co-design approach. Initially, we show that existing ADC-Less partial sum quantization (binary or ternary) algorithms for analog CiM accelerators rely heavily on numerous floating-point scale factors for accurate computations. However, reducing the number of scale factors often results in lower accuracy. Moreover, processing these scale factors would require complex hardware. To overcome this, we quantize the scale factors during training. Next, we propose \papername, which consists of an analog CiM crossbar to perform MVM operation and a digital CiM array to process these scale factors efficiently. We also introduce a novel method for performing in-memory subtraction and addition, a crucial step in processing scale factors. Furthermore, by exploiting sparsity in ternary quantization, we achieve an additional 15\% energy savings compared to binary quantization through clock-gating the column peripherals in the DCiM array. Our system-level evaluation using a cycle-accurate simulator shows up to 28$\times$ and 12$\times$ reduction in energy compared to the baseline that uses 7-bit and 4-bit ADCs, respectively. 

\section*{Acknowledgments}
This work was supported by the Center for the Co-Design of Cognitive Systems (COCOSYS), a
DARPA-sponsored JUMP center of Semiconductor Research Corporation (SRC). The authors would also like to thank Trishit Dutta for assisting with debugging errors in the Synopsis Design Compiler.

{\small
\bibliographystyle{ieee_fullname}
\bibliography{main}
}

\end{document}